\documentclass{article}
\usepackage{amssymb}\usepackage{amsmath}\usepackage{epsfig}
\begin{document}
\title{Nonlinear supratransmission as \\ a fundamental instability}
\author{ J\'er\^ome LEON\\
Physique Math\'ematique et Th\'eorique, CNRS-UMR 5825\\
34095 MONTPELLIER  (FRANCE)}
\date{}\maketitle

\begin{abstract} The {\em nonlinear supratransmission} is the property of a
nonlinear system possessing a natural forbidden band gap to transmit energy of
a signal with a frequency in the gap by means of generation of nonlinear modes
(gap solitons). This process is shown to result from a generic instability of
the evanescent wave profile generated in a nonlinear medium by the incident
signal.\end{abstract}

\section{Introduction}

A plane wave scattering onto a medium which possess a natural forbidden band
gap, is totally reflected if its frequency  falls in the gap. Such may not be
the case if the medium is nonlinear, as  the sine-Gordon model considered in
\cite{jg-al} where the incident wave was shown to generate, above some
amplitude threshold, a sequence of nonlinear modes propagating in the medium.

This threshold was given an analytic expression in \cite{prl} (still in the
sine-Gordon case) and the property of the nonlinear medium to transmit energy
in the gap has been called {\em nonlinear supratransmission}. It has been shown
then to hold for different models (nonlinear Klein-Gordon, double sine-Gordon,
Josephson transmission lines), and has been experimentally realized on a chain
of coupled pendula \cite{jphysc}.

Energy transmission by gap soliton generation is a phenomenon that has been
experimentally discovered in Bragg media with Kerr nonlinearity
\cite{taverner}.  In that case the underlying simplified model is the {\em
coupled mode system} \cite{desterke} and it has been recently demonstrated that
the very mechanism of the switch from a Bragg mirror to a medium with high
transmissivity is the nonlinear supratransmission \cite{bragg}. Again there the
theory allows for an analytic expression of the threshold amplitude, based on
the expression of the soliton solution of the coupled mode system found in
\cite{aceves}.

We demonstrate here that the process of nonlinear supratransmission is a
natural consequence of a generic instability of the evanescent profile
generated in a nonlinear medium when it is constrained at one end by a periodic
boundary driving of frequency in a forbidden band gap.

\section{General context}

Let us consider the Lagrangian of a scalar field $u(x,t)$ in the half-plane
$(x,t)\in {\mathbb R}^+\times{\mathbb R}$
\begin{equation}
{\cal L}=\int_0^\infty dx\,L(u,u_t,u_x)\ ,\quad 
L=\frac12u_t^2-\frac12u_x^2-\phi(u)\ .\end{equation}
The equation of motion $\delta{\cal L}=0$ reads
\begin{equation}\label{wave-eq}
x> 0\ :\ u_{tt}-u_{xx}+\phi'(u)=0\ ,\end{equation}
where {\em prime} denotes differentiation with respect to the argument.
The wave equation \eqref{wave-eq} is submitted to Dirichlet boundary conditions 
\begin{equation}\label{dirichlet}
u(0,t)=f(t)\ ,\quad u(\infty,t)=0\ ,\end{equation}
associated with Cauchy initial data $u(x,0)$ and $u_t(x,0)$, which can be
vanishing or adapted to the boundary forcing. Note that this approach 
can be extended to Neumann boundary conditions.

 From now on we adopt the  quite generic assumtion of a  potential energy
$\phi(u)$  which  possess a local minimum locally symmetric. By a convenient
choice of the reference frame, this minimum can be taken to be located at
$(\phi,u)=(0,0)$. Then the symmetric requirement means that the potential is
{\em locally} an even function of $u$. Moreover, by a convenient choice of the
units of $x$ and $t$ we can always scale the value of $\phi''(0)$ to the value
$1$. In summary the potential $\phi(u)$ obeys the following requirements
\begin{equation}\label{constr}
\phi(0)=0\ ,\quad \phi'(0)=0\ ,\quad \phi''(0)=1\ ,\quad \phi'''(0)=0\ .
\end{equation}
Note that the {\em local parity assumption} $\phi'''(0)=0$ is not essential but
simplifies the expressions.

As a consequence the equation \eqref{wave-eq},
by Taylor expansion of $\phi'(u)$ about the minimum $u=0$, has the linear limit
\begin{equation}\label{lin-lim}
 u_{tt}-u_{xx}+u=0\ ,\end{equation}
with dispersion relation
\begin{equation}\label{disp-rel}\omega^2=1+k^2\ ,\end{equation}
having the gap $[-1,+1]$ in the frequency.

The second restiction is on the class of boundary datum $f(t)$ which we take
to be a periodic forcing with a frequency belonging to the forbidden band gap.
More precisely we assume
\begin{equation}\label{periodic-bound}
u(0,t)=A\,\cos(\Omega t)\ ,\quad |\Omega|<1\ .\end{equation}
This problem is quite generic and describes e.g. the scattering of a 
monochromatic wave into a medium in one of its forbidden band gaps. 

 For the linear limit \eqref{lin-lim}, the boundary condition 
\eqref{periodic-bound} would generate the solution ($x>0$)
\begin{equation}\label{evanescent}
u_0(x,t)=A\,\cos(\Omega t)\,e^{-Kx}\ ,\quad \Omega^2=1-K^2\ ,\end{equation}
called the linear evanescent wave. This solution is a reasonable approximate 
solution of the nonlinear evolution as long as the amplitude $A$ is small 
enough.

We show hereafter that in the nonlinear case, under some additional simple
conditions on the potential $\phi(u)$, the medium can present {\em nonlinear
supratransmission} as a result of a fundamental instability generated  by  the
evanescent wave, above some threshold for the amplitude $A$. 

\section{Perturbation scheme}

Let us start with the simplest nonlinear correction resulting from the
assumption \eqref{constr}, namely the nonlinear wave equation
\begin{equation}\label{nonlin-lim}
u_{tt}-u_{xx}+u+\frac16\alpha\, u^3=0\ ,\end{equation}
where the nonlinear factor $\alpha$ is defined by
\begin{equation}\label{alpha-def}
\alpha=\left.\frac{\partial^4 \phi}{\partial u^4}\right|_{u=0}\ .
\end{equation}
\begin{figure}[ht]
\centerline{\epsfig{file=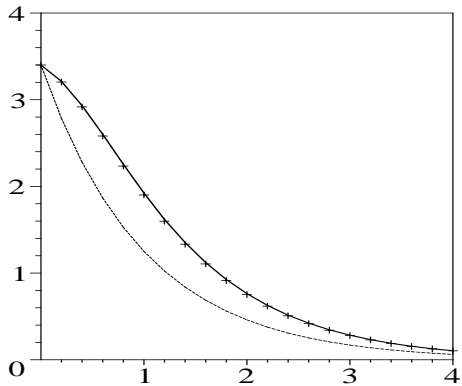,height=5cm,width=6cm}}
\caption{\label{fig:profiles}
Plot of the actual nonlinear evanescent profile $u_1(x)$
(full line) compared to the linear one $u_0(x)$ (dashed line) in the 
static case $\Omega=0$ for $A=3.4$. The dots are the results of a numerical
simulation of \eqref{nonlin-lim}.}
\end{figure}

It is essential to remark at this step that the first effect of the
nonlinearity on a boundary driving \eqref{periodic-bound} is to {\em deform}
the (linear) evanescent profile \eqref{evanescent}.  Such a deformation can be
illustrated analytically and numerically in the static case $\Omega=0$ in
\eqref{periodic-bound}. Indeed  the equation \eqref{nonlin-lim} with
$\alpha=-1$ for $u(x)$ submitted to $u(0)=A$, possess the solution
\begin{equation}\label{deformed}
u_1(x)=\frac{2\sqrt{3}}{\cosh(x-x_0)}\ ,\quad 
x_0=-|{\rm arccosh}(\frac{2\sqrt{3}}{A})|\ .\end{equation}
This is the profile displayed on the figure \ref{fig:profiles} with $A=3.4$
(full line) in perfect agreement to numerical simulations
(dots). The numerical method is explained in the appendix and here we have
solved \eqref{nonlin-lim} with a boundary datum
$u(0,t)$ slowly increasing from $0$ to the value $A=3.4$ and the solution
(represented by the crosses in fig. \ref{fig:profiles}) is plotted long
after the boundary datum has been setteled.
We have plotted also for comparaison to the exponential (linear)
shape \eqref{evanescent} namely $u_0(x)=Ae^{-x}$ (dashed line).

In the dynamical case ($\Omega\ne0$) we  denote by $u_1(x,t)$ the {\em deformed
evanescent profile}, solution of \eqref{nonlin-lim}, hence required to obey
the boundary value \eqref{periodic-bound}, namely 
\begin{equation}u_1(0,t)=A\,\cos(\Omega t)\ .\end{equation}
Although this solution $u_1(x,t)$ may not be known analytically, we shall
actually discover that the worst approximation of it, namely the linear
evanescent profile $u_0(x,t)$ itself, still generates the seeked instability,
hence providing a qualitative understanding of the generation of nonlinear
supratransmission.

\section{Instability criterion}

The second effect of the boundary driving is obtained
by studying the perturbation of the exact solution $u_1(x,t)$ of 
\eqref{nonlin-lim}  as
\begin{equation}u(x,t)=u_1(x,t)+\epsilon\eta(x,t)\ ,\quad \epsilon\ll1\ .
\end{equation}
At order $\epsilon$, the evolution for the perturbation $\eta$ then reads
\begin{equation}\label{perurb-eq}
x>0\ ,\quad \eta_{tt}-\eta_{xx}+\eta[1+V]=0\ ,\quad
V(x,t)=\frac12\alpha u_1^2(x,t)\ .\end{equation}
This linear wave equation in the potential $V$ on the semi-axis $x>0$
is our  main tool and contains all the ingredients required for a
qualitative description of the onset
of instability.

The linear equation \eqref{perurb-eq} can be thought of as a two-dimensional
scattering problem on the half-plane for a potential which is periodic in the
infinite $t$-dimension and localized around $x=0$ in the positive
$x$-direction. The study of this mixed periodic-localized 2D scattering problem
is a tough technical question that we report in a future investigation.

However we show hereafter that the instability can be qualitatively understood
by considering the {\em static case} where $V$ is assumed to depend only on
the space variable $x$. The scattering problem can then be reduced to the 
radial S-wave  Schr\"odinger spectral problem 
\begin{equation}\label{sturm}
\psi_{xx}(\zeta,x)+\left[(\lambda^2-1)-V(x)\right]\psi(\zeta,x)=0\ ,
\end{equation}
for the spectral parameter $\zeta$ and eigenfunction $\psi(\zeta,x)$ defined by
\begin{equation}\label{def-psi}
\zeta^2=\lambda^2-1\ ,\quad \eta(x,t)=e^{i\lambda t}\psi(\zeta,x)\ .
\end{equation}
We argue then that equation \eqref{perurb-eq} develops an instability if the
potential $V$  possess a bound state, say $\zeta=ip$, with {\em sufficient 
energy}, namely if
\begin{equation}\label{criterion}
\zeta^2=-p^2\ ,\quad p>1\ \Rightarrow\ \lambda^2<0\ ,\end{equation}
which causes the first order correction $\eta(x,t)$ to diverge in time.
In particular it is clear from \eqref{perurb-eq} that the instability will
develop only when $\alpha<0$ as a necessary (not sufficient) condition to have
a bound state for $V$. In order to allow for a comparaison with sine-Gordon,
and because the value of $\alpha$ in \eqref{nonlin-lim} can be scaled with $A$,
we set
\begin{equation} \alpha=-1\ .\end{equation}

\begin{figure}[ht]
\centerline{\epsfig{file=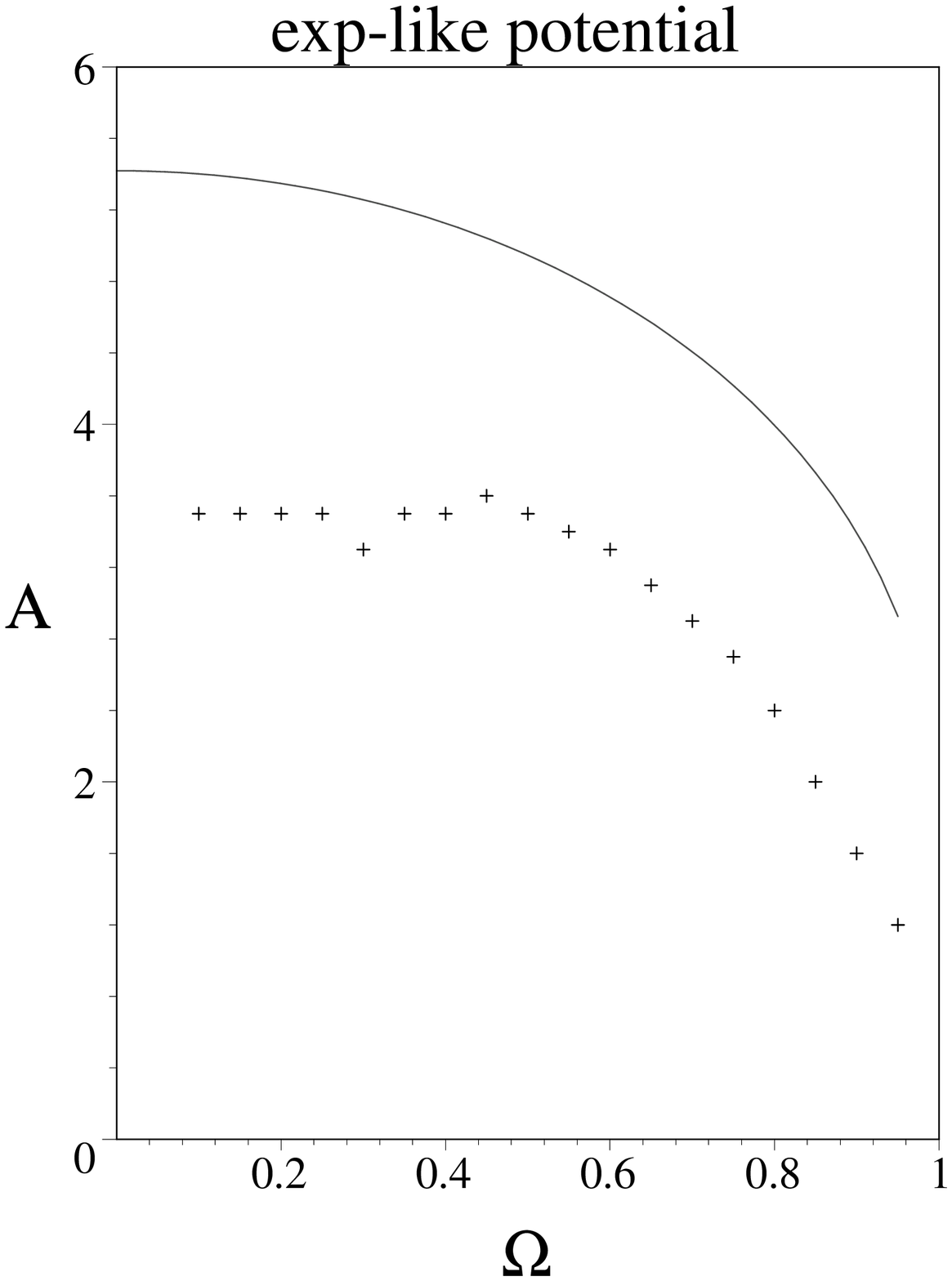,height=5cm,width=5cm}
\epsfig{file=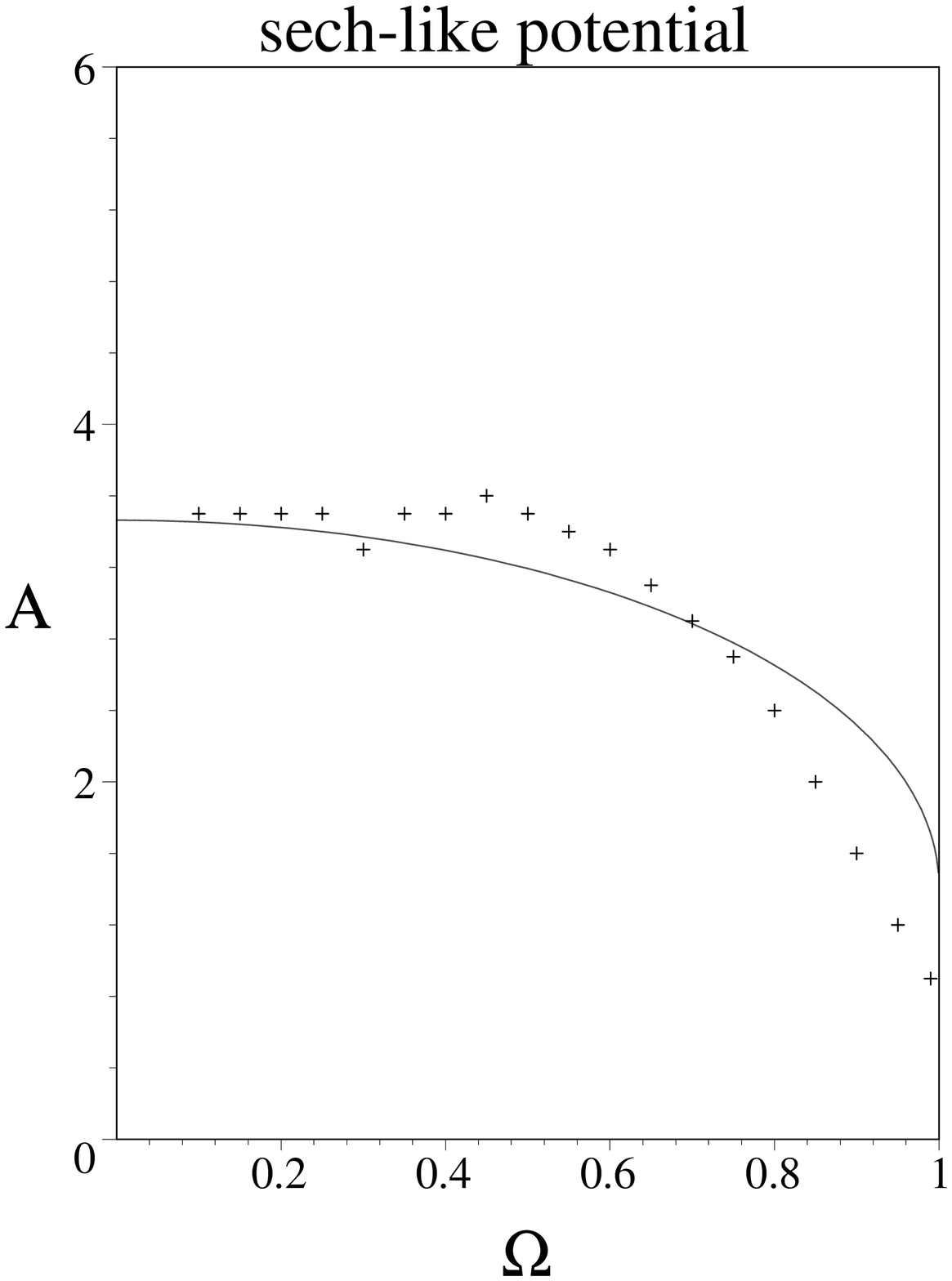,height=5cm,width=5cm}}
\caption{
Comparaison of the threshold of instability for the full
equation \eqref{nonlin-lim} (crosses) with the theory obtained from 
\eqref{perurb-eq}  with   $\alpha=-1$ for $u_1=A\exp[-Kx]$ (full line
left figure) and $u_1=A/\cosh(Kx)$ (full line right figure).
\label{fig:compare-seuil}}
\end{figure}

The potential $V(x)$ is obtained from the {\em deformed evanescent profile}
$u_1(x,t)$ by considering its extremum on a period, namely we take
\begin{equation}\label{max}
V(x)=-\frac1 2\max_{t}\{ u_1^2(x,t)\}\ .\end{equation}
We now illustrate the obtained expression for the threshold of instability for
two particular choices for $u_1$, first the linear evanescent profile
\eqref{evanescent}, second a standard sech-type profile. Both cases allow for
an explicit computation of this threshold and we compare in figure
\ref{fig:compare-seuil} these results to numerical simulation of the PDE
\eqref{nonlin-lim}.

\section{The exponential potential}

The choice $A\exp[-Kx]$ for $\max\{ u_1^2(x,t)\}$ provides from \eqref{sturm}
the following spectral problem on the half-line
\begin{equation}\label{sturm-exp}x>0\ ,\quad
\psi_{xx}(\zeta,x)+\left[\zeta^2+\frac12A^2e^{-2Kx}\right]\psi(\zeta,x)=0\ .
\end{equation}
The two (positve, real-valued) parameters are the amplitude $A$ and the 
{\em wave number} $K=\sqrt{1-\Omega^2}$.

The direct scattering problem \cite{newton,chad-sab} essentially consists in
finding the Jost solution of \eqref{sturm-exp} which can be written
\begin{equation}\label{jost-exp}
f(\zeta,x)=e^{i\zeta x}\,{}_0F_1(1-i\frac\zeta K;-\frac{A^2}{8K^2} e^{-2Kx})\ .
\end{equation}
This solution is usually written in terms of the Bessel function of the first
kind of order $-i\zeta/K$. We find it more convenient to express it in terms
of the generalized hypergeometric function ${}_0F_1$. Indeed it readily
provides the asymptotic behavior
\begin{equation}\label{asymp}
f(\zeta,x)e^{-i\zeta x}\underset{x\to\infty}{\longrightarrow}\ 1\ ,
\end{equation}
and the Jost function
\begin{equation}\label{func-exp}
f(\zeta,0)={}_0F_1(1-i\frac\zeta K;-\frac{A^2}{8K^2})\ .
\end{equation}

The bound states are the zeroes of this Jost function in the upper complex
$\zeta$-plane, actually located on the imaginary axis as the potential is real
valued. Our instability criterion \eqref{criterion} then tells that an
instability for $\eta(x,t)$ will develops as soon as the amplitude $A$ reaches
the value $A_s$ for which the first bound state reaches the value $\zeta=i$.
This occurs for each value of the parameter $K$ and thus the threshold $A_s(K)$
is defined by the implicit equation
\begin{equation}
{}_0F_1(1+\frac1 K;-\frac{A_s^2}{8K^2})=0\ .\end{equation}
We have solved this equation numerically and the result is plotted in terms
of $\Omega=\sqrt{1-K^2}$ in the left figure \ref{fig:compare-seuil}. 

It is compared there with the results of the numerical solution (see appendix)
of \eqref{nonlin-lim} (with $\alpha=-1$) submitted to the boundary driving
\eqref{periodic-bound}, namely $u(0,t)=A\cos\Omega t$. Although the theoretical
threshold does not fit perfectly the simulations, the principle of the
nonlinear supratransmission is qualitatively understood. The discrepancy
results both from the asumption of an exponential profile for $u_1(x,t)$ and a
static approximation of it.

\section{The Eckart potential}

A better fit is obtained by assuming instead of the exponential profile the
following sech-type profile obtained from the solution \eqref{deformed} at its
maximum amplitude (i.e. for $x_0=0$ where the instability is expected)
\begin{equation}
u_1=\frac A{\cosh Kx}\ ,\quad K=\sqrt{1-\Omega^2}\ .\end{equation}
The presence of a periodic boundary datum is taken into account by the
parameter $K$ (wave number).  This choice is expected to give good results for
small $\Omega$ and it leads to the spectral problem 
\begin{equation}\label{sturm-sech}x>0\ ,\quad
\psi_{xx}(\zeta,x)+\left[\zeta^2+\frac12
\frac{A^2}{\cosh^2Kx}\right]\psi(\zeta,x)=0\ .
\end{equation}
This potential belongs to the Eckart class \cite{newton} and its Jost solution 
reads
\begin{equation}\label{jost-sech}
f(\zeta,x)=e^{i\zeta x}\left(1+e^{-2Kx}\right)^\beta\,
{}_2F_1(\beta,\beta-i\frac\zeta K;1-i\frac\zeta K;-e^{-2Kx})\ ,
\end{equation}
with the parameter
\begin{equation}\label{beta}
\beta=\frac12\left(1-\sqrt{1+2A^2/K^2}\right)\ .\end{equation}
This solution does obey the asymptotic \eqref{asymp} and the Jost function is
\begin{equation}\label{func-sech}
f(\zeta,0)=2^\beta\,
{}_2F_1(\beta,\beta-i\frac\zeta K;1-i\frac\zeta K;-1)\ .
\end{equation}

According to the instability criterion \eqref{criterion} we then seek the value
$A_s$ for the threshold amplitude generating a bound state (zero of this Jost
function on the upper imaginary $\zeta$-axis) at the value $\zeta=i$. From
\cite{bateman} we can write also \eqref{func-sech} as
\begin{equation}\label{func-spec}
f(\zeta,0)=\frac{2^{i\zeta/K}\sqrt{\pi}\,\Gamma(1-i\frac\zeta K)}
{\Gamma(\frac{1+\beta}2-i\frac\zeta{2K})\,
\Gamma(1-\frac{\beta}2-i\frac\zeta{2K})}\ ,\end{equation}
where $\Gamma(v)$ is the standard Euler Gamma function that possess
poles at $v=0,-1,\cdots$. 

The bound states are thus given by the poles of the
function $\Gamma(\frac{1+\beta}2-i\frac\zeta{2K})$ on the positive imaginary
$\zeta$-axis, and the first bound state reaches the point $\zeta=i$  as soon
as $\beta$ verifies
\begin{equation}
\frac{1+\beta}2+\frac1{2K}=0\ .\end{equation}
Replacing there the value \eqref{beta} of  the parameter $\beta$, the
threshold of instability $A_s$ can then be expressed in terms of
$\Omega=\sqrt{1-K^2}$ as
\begin{equation}\label{threshold-sech}
A_s^2=4(1-\Omega^2)+6\sqrt{1-\Omega^2}+2\ .\end{equation}
This is the curve represented in the right graph of figure 
\ref{fig:compare-seuil}. We see then that the sech-like profile provides
as expected a  much better fit of the  instability threshold, especially
for small $\Omega$ where the static approximation becomes more accurate.

\section{The sine-Gordon case}

So far we have demonstrated the existence of an onset of instability in the
nonlinear simple model \eqref{nonlin-lim} (with $\alpha<0$). In that
case the instability causes the solution to diverge because the related
potential $\phi(u)=u^2/2-u^4/24$ is not confining for large $u$. 

To illustrate the nonlinear supratransmission as an effect of the instability,
we consider now the sine-Gordon case $\phi(u)=1-\cos(u)$ treated in
\cite{prl,jphysc}, namely
\begin{equation}\label{SG}x>0\ :\quad
u_{tt}-u_{xx}+\sin u=0\ ,\quad u(0,t)=A\cos\Omega t\ .\end{equation}
In this situation, the solution induced by the boundary datum of amplitude
{\em below} the threshold of instability is the {\em non-travelling} breather 
\begin{equation}\label{breather}
u_b(x,t)=4\arctan\left[\frac K\Omega
\frac{\cos(\Omega t)}{\cosh(K(x-x_b)}\right]\ ,\quad x_b<0\ ,
\end{equation}
where $x_b$ is the breather center. 

The onset of nonlinear supratransmission is reached, at given frequency
$\Omega$, for the maximum amplitude of $u_b(0,t)$, obtained of course
for $x_b=0$. Let $u_2(x,t)$ denote the solution $u_b$ for $x_b=0$, namely
\begin{equation}\label{max-breath}
u_2(x,t)=4\arctan\left[\frac K\Omega
\frac{\cos(\Omega t)}{\cosh(Kx}\right]\ .
\end{equation}
The first order perturbation $\epsilon\eta(x,t)$ about the solution $u_2$
obeys from \eqref{SG} the linear evolution
\begin{equation}\label{perurb-sg}
x>0\ ,\quad \eta_{tt}-\eta_{xx}+\eta[1+V]=0\ ,\quad
V(x,t)=\cos (u_2)-1\ .\end{equation}
that replaces equation \eqref{perurb-eq} used precedingly. The potential $V$
can also be written
\begin{equation}\label{pot-breath}
V(x,t)=-8\frac{K^2\cos^2\Omega t}{\Omega^2\cosh^2Kx}
\left[1+\frac{K^2\cos^2\Omega t}{\Omega^2\cosh^2Kx}\right]^{-2}\ .
\end{equation}

Our purpose here is  to show that the threshold of instability of the
linear evolution \eqref{perurb-sg} with the {\em stationary approximation}
\eqref{max}, namely
\begin{equation}\label{static}
V(x)=-8\frac{B^2}{\cosh^2Kx}
\left[1+\frac{B^2}{\cosh^2Kx}\right]^{-2}\ .
\end{equation}
matches reasonably the theoretical prediction, namely the value $B_s$ of
the parameter $B$ given from \eqref{pot-breath} by
\begin{equation}\label{Bs}
B_s= \frac K\Omega = \frac{\sqrt{1-\Omega^2}}\Omega\ .\end{equation}
\begin{figure}[ht]
\centerline{\epsfig{file=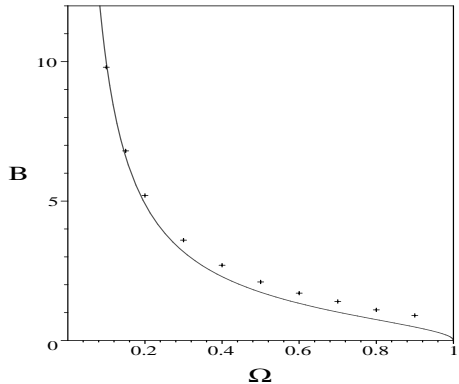,height=5cm,width=6cm}}
\caption{Onset of instability for the parameter $B$ of $V(x)$ 
given by numerical simulations of \eqref{perurb-sg}  with
\eqref{static} (dots) compared to the prediction \eqref{Bs}.\label{fig:SG}}
\end{figure}
The drawback here is that the potential \eqref{static} does not lead to an
explicitely solvable spectral problem \eqref{sturm}.  This is why we have used
numerical simulations and shown  by figure \ref{fig:SG} that the stationary
limit does produce a consistent result: the threshold of instability of the
PDE \eqref{perurb-sg} obtained by varying $B$, at each fixed $\Omega$, fits
suprisingly well the exact prediction \eqref{Bs}.

Note finally that in the sine-Gordon case we have an analytic expression of the
threshold amplitude that generates nonlinear supratransmission \cite{prl},
namely
\begin{equation}\label{threshold-sg}
A_s(\Omega)=4\arctan[\sqrt{1-\Omega^2}/\Omega]\ .\end{equation}
This is simply the maximum amplitude of the breather \eqref{max-breath}, which
is indeed reached when $B$ in \eqref{static} takes the value \eqref{Bs}.
Above this value the instability comes into play and generates a nonlinear 
mode that propagates in the medium carrying energy, what have been called
nonlinear supratransmission and have been experimentally realized on a chain
of pendula \cite{jphysc}.

\section{Conclusion}

We have demonstrated that a nonlinear medium, under some generic conditions on
the structure of the nonlinearity (existence of a forbidden band gap, local
minimum for the potential, ...) may create an instability of the evanes\-cent
profile generated by a periodic boundary driving at a frequency in the gap.
This instability is the generating process at the origin of nonlinear
supratransmission, when the nonlinear medium supports nonlinear modes
(breathers, solitons, kinks, ...).

The instability has been described by using a stationary limit for the
evanescent profile which has allowed us to prove that the occurence of the
instability is related to the appearence of a bound state with sufficient
energy, according to \eqref{criterion}. 

We have dealt here with a scalar wave-like equation and it is worth
mentionning that the process works also for different types of systems
like for instance the coupled mode equations in Bragg media \cite{bragg}.
The generating  instability in this case will be considered in a future work.

\appendix\section{Numerical simulations}

In order to integrate numerically the partial differential equations to deal
with, we have used two different approaches depending wether the PDE is linear
or not. 

In the nonlinear case, such as equation \eqref{wave-eq}, we have
used an explicit second order scheme in space and solved for time evolution as
a set of coupled differential equations. More precisely we have replaced
\eqref{wave-eq} with the set of ODE's (overdot means time differentiation)
\begin{equation}
\ddot u_n-\frac1{h^2}(u_{n+1}-2u_n+u_{n-1})+\phi'(u_n)=0\ ,\quad n=0..N\ ,
\end{equation} 
where $h$ is the chosen grid step and $u_n(t)=u(nh,t)$. 

The boundary datum \eqref{periodic-bound} is smoothly setteled by assuming
\begin{equation}\label{bound-smooth}
u_0(t)=A \frac12\left[1+\tanh(p(t-t_0))\right]\cos\Omega t\ ,\end{equation}
where $p$ and $t_0$ are the parameters to select. A semi-infinite line is
simulated by an absorbing end consisting of a damping term $\gamma \dot u_n$
included in the equation where the intensity $\gamma$ slowly varies from $0$ to
$1$ over the last $40$ cells.  

The results presented in figures \ref{fig:profiles} and \ref{fig:compare-seuil}
are obtained with $N=100$ spatial mesh points, $h=0.2$ grid spacing over a time
of integration $t_m=200$. The parameters for the boundary field are $p=0.2$ and
$t_0=20$.  The resulting set of coupled ODEs is then solved through the
subroutine {\tt dsolve} of the {\tt MAPLE8} software package that uses a
Fehlberg fourth/fifth order Runge-Kutta method.

In the linear case, namely for equation \eqref{perurb-sg}, we have used the
{\tt pdsolve} routine of  {\tt MAPLE8} over a length $L=10$ and a space-time
step of $0.1$. It uses a centered implicit scheme, of second order in space and
time, with a first order boundary condition. To test the stability of an
equation like \eqref{perurb-sg} we have solved it for the following set of
initial-boundary values \begin{equation} \eta(0,t)=0.01\ ,\quad\eta(L,t)=0.01\
,\quad\eta(x,0)=0.01\ ,\quad \eta_t(x,0)=0\ .\end{equation} In the stable
region we observe small amplitude oscillations that grow when the potential
parameter get close to the instability threshold. Above this threshold the
instability manisfests as an exponential growth of the solution.  The points
that result in the figure \ref{fig:SG} denote the first value of the parameter
$B$ for which instability settles in, with an absolute precision of $10^{-1}$.

\end{document}